\documentstyle[preprint,aps,psfig]{revtex}

\begin{document}

\draft
\tighten
\preprint{
\vbox{
\hbox{February 1996}
\hbox{DOE/ER/40762-080}  
\hbox{U.MD. PP\#96-067}  
\hbox{ADP--96--9/T214}
}}

\title{Neutron Structure Functions From Nuclear Data
\footnote{Invited talk presented by W. Melnitchouk at the
{\em Cracow Epiphany Conference on Proton Structure},
5-6 Jan. 1996, Cracow, Poland.  To appear in {\em Acta Phys. Pol.}}}     

\author{W.Melnitchouk}
\address{Department of Physics, 
         University of Maryland, 
         College Park, MD 20742}
\author{A.W.Thomas}
\address{Department of Physics and Mathematical Physics,
         University of Adelaide,
         5005, Australia.}

\maketitle

\begin{abstract}
The spin-averaged structure function of the neutron,     
$F_2^n$, is extracted from recent deuteron data,
taking into account the small but significant corrections 
from nuclear effects in the deuteron.
At small $x$, the $F_2^D/F_2^p$ ratio measured by the 
New Muon and Fermilab E665 Collaborations is interpreted 
to suggest a small amount of shadowing in deuterium, 
which acts to enhance $F_2^n$ for $x \alt 0.1$. 
A careful treatment of Fermi motion, binding and nucleon 
off-shell effects in the deuteron also indicates that the 
neutron/proton structure functio ratio as $x \rightarrow 1$
is consistent with the perturbative QCD expectation of 3/7, 
but larger than the traditional value of 1/4.  
\end{abstract}

\section{Introduction}

The quark structure of the nucleon is one of the most fundamental
aspects of hadron physics. Deep inelastic scattering (DIS) of 
leptons from hydrogen has yielded a wealth of information on 
the quark and gluon substructure of the proton.
The absence of free neutron targets means, however, that it is
difficult to obtain direct data on $F_2^n$.
As a result, one usually uses deuterium targets, and extracts
neutron structure information from a knowledge of the proton
structure function, and the nucleon wave function in the deuteron.  
The accuracy of the extracted neutron data naturally depends on 
the level of understanding of the nuclear physics in the deuteron, 
as well as on the extraction procedure itself.
Both of these issues are carefully addressed in this paper.

The treatment of nuclear effects is divided into two regions: 
large $x$ ($x \agt 0.3$) and small $x$ ($x \alt 0.2$).
In the context of the multiple scattering framework, the 
large-$x$ effects are described within the impulse approximation,
in which the virtual photon interacts with only one nucleon in the 
deuteron, while the other nucleon 
remains spectator to the interaction.  
The impulse approximation provides a natural framework within 
which effects from nuclear binding, Fermi motion, and nucleon 
off-shellness can be incorporated.  
At small $x$, on the other hand, there are important contributions 
from the rescattering of the probe from both nucleons in the 
deuteron, which gives rise to the phenomenon known as nuclear 
shadowing.

\section{Large $x$}

Away from the small-$x$ region ($x \agt 0.3$), the dominant
contribution to the deuteron structure function can be computed 
from the impulse approximation.
Here the total $\gamma^* D$ amplitude is factorized into 
$\gamma^* N$ and $N D$ amplitudes, although, contrary to what 
is often assumed, this factorization does not automatically lead 
to a factorization of cross sections in the convolution model
\cite{CONV}.

\subsection{Binding, Fermi Motion and Off-Shell Effects}

Starting from the impulse approximation, one can show that in
the non-relativistic approximation the nuclear structure 
function can be written in convolution form, in which the 
structure function of the nucleon is smeared with a momentum
distribution, $f_{N/D}(y)$, 
of nucleons in the deuteron \cite{CONV,KMPW}:
\begin{eqnarray}
\label{con}
F_2^{D\ {\rm (conv)}}(x,Q^2)
&=& \int dy\ f_{N/D}(y)\
    F_2^N\left( {x \over y},Q^2 \right),
\end{eqnarray}
where $F_2^N = F_2^p + F_2^n$.
Equation (\ref{con}) is correct to order $(v/c)^2$ (with $v$ the nucleon
velocity), provided we also neglect the possible $p^2$ dependence
in the nucleon structure function. 
The distribution function $f_{N/D}(y)$ is determined by the 
deuteron wave functions $u, w, v_t$ and $v_s$ \cite{DREL}
(corresponding to the deuteron's $S, D$ and triplet and 
singlet $P$ waves) \cite{MSTD}:   
\begin{eqnarray}
f_{N/D}(y)
&=& { M_D \over 4 }\ y\
    \int_{-\infty}^{p^2_{max}} dp^2\
    {E_p \over p_0}\ 
\left( u^2({\bf p}^2)\ +\ w^2({\bf p}^2)\
    +\ v_t^2({\bf p}^2)\ +\ v_s^2({\bf p}^2)
\right) 
\end{eqnarray}
where $p$ is the interacting nucleon's four-momentum, 
with a maximum squared value  
$p^2_{max} = y M_D^2 - y M^2/(1-y)$,
and $E_p\ =\ \sqrt{M^2 + {\bf p}^2}$.

As explained in Refs.\cite{MSTD,MST}, explicit corrections to 
Eq.(\ref{con}), which cannot be written in convolution form,
arise when the bound nucleons' off-mass-shell structure is 
taken into account:
\begin{eqnarray}
F_2^D(x,Q^2)
&=& F_2^{D\ {\rm (conv)}}(x,Q^2)\
 +\ \delta^{\rm (off)} F_2^D(x,Q^2).
\end{eqnarray}
The correction $\delta^{\rm (off)} F_2^D$ receives contributions
from the off-shell components in the deuteron wave function, as 
well as from the off-mass-shell dependence of the bound nucleon 
structure function \cite{MSTD} (i.e. in the $p^2 \rightarrow M^2$ 
limit, in which the $P$-state wave functions also vanish, one has 
$\delta^{\rm (off)} F_2^D \rightarrow 0$).

\begin{figure}
\centering{\ \psfig{figure=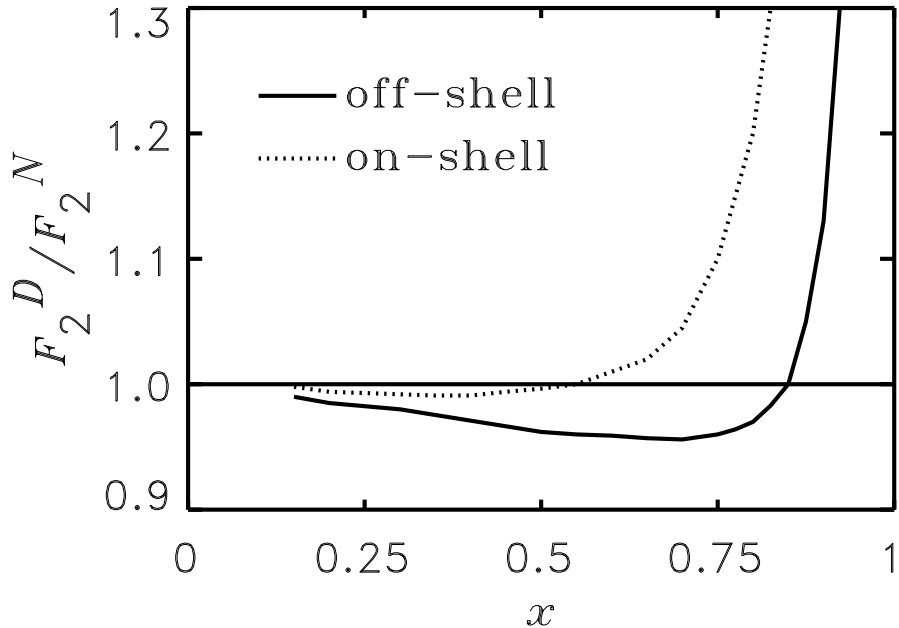,height=10cm}}
\caption{$F_2^D/F_2^N$ ratio as a function of $x$ for the
         off-shell model of Refs.\protect\cite{MSTD,MST} (solid)
         and the on-shell model of
         Ref.\protect\cite{FS78} (dotted).}
\end{figure}

In Refs.\cite{MSTD,MST} the structure function $F_2^N$ was
modeled in terms of relativistic quark--nucleon vertex functions,
which were parametrized by comparing with available data for the parton 
distribution functions.
The off-shell extrapolation of the $\gamma^* N$ interaction 
was modeled assuming no additional dynamical $p^2$ dependence 
in the quark--nucleon vertices.  
This enabled an estimate of the correction 
$\delta^{\rm (off)} F_2^D$ to be made, which was found to be
quite small, of the order $\sim 1-2\%$ for $x \alt 0.9$.
The result of the fully off-shell calculation from Ref.\cite{MSTD}
is shown in Fig.1 (solid curve), where the ratio of the total
deuteron to nucleon structure functions ($F_2^D/F_2^N$) is plotted.
Shown also is the result of an on-mass-shell calculation from
Ref.\cite{FS78} (dotted curve), which has been used in many
previous analyses of the deuteron data \cite{WHIT,EMC}.
The most striking difference between the curves is the fact
that the on-shell ratio has a very much smaller trough at
$x \approx 0.3$, and rises faster above unity (at $x \approx 0.5$) 
than the off-shell curve, which has a deeper trough, at 
$x \approx 0.6-0.7$, and rises above unity somewhat later 
(at $x \approx 0.8$).

The behavior of the off-shell curve in Fig.1 is qualitatively
similar to that found by Uchiyama and Saito \cite{US}, Kaptari 
and Umnikov \cite{KU}, and Braun and Tokarev \cite{BT}, who also 
used off-mass-shell kinematics, but did not include the (small)
non-convolution correction term $\delta^{\rm (off)} F_2^D$.
The on-shell calculation \cite{FS78}, on the other hand, was 
performed in the infinite momentum frame where the nucleons 
are on their mass shells and the physical structure functions 
can be used in Eq.(\ref{con}).
One problem with this approach is that the deuteron wave functions 
in the infinite momentum frame are not explicitly known.
In practice one usually makes use of the ordinary non-relativistic
$S$- and $D$-state deuteron wave functions calculated in the 
deuteron rest frame, a procedure which is analogous to including 
only Fermi motion effects in the deuteron.
In addition, one knows that the effect of binding in the infinite
momentum frame shows up in the presence of additional Fock 
components (e.g. $NN$-meson(s) ) in the nuclear wave function, 
which have not yet been computed.

Clearly, a smaller $D/N$ ratio at large $x$, as in the off-shell
calculation, implies a larger neutron structure function in this
region.
To estimate the size of the effect on the $n/p$ ratio requires one
to ``deconvolute'' Eq.(\ref{con}) in order to extract $F_2^n$.

\subsection{Extraction of $F_2^n$}

%
To study nuclear effects on the neutron structure function arising
from different models of the deuteron, one must eliminate any effects
that may arise from the extraction method itself.
We therefore use exactly the same extraction procedure as used 
in previous SLAC \cite{WHIT} and EMC \cite{EMC} data analyses,
namely the smearing (or deconvolution) method discussed by Bodek 
{\em et al.} \cite{BODEK}.
This method involves the direct use of the proton and deuteron 
{\em data}, without making any assumption for $F_2^n$ itself.
For completeness let us briefly outline the main ingredients in 
this method.
(For alternative methods of unfolding the neutron structure 
function see for example Refs.\cite{UKK,VATO}.)

Firstly, one subtracts from the deuteron data, $F_2^D$, the 
additive, off-shell corrections, $\delta^{\rm (off)} F_2^D$, 
to give the convolution part, $F_2^{D\ {\rm (conv)}}$.
Then one smears the proton data, $F_2^p$, with the nucleon 
momentum distribution function $f_{N/D}(y)$ in Eq.(\ref{con}) 
to give $\widetilde{F}_2^p \equiv F_2^p/S_p$.
The smeared neutron structure function, $\widetilde{F}_2^n$, 
is then obtained from
\begin{eqnarray}
\label{F2nsm}
\widetilde{F}_2^n &=& F_2^{D\ {\rm (conv)}} - \widetilde{F}_2^p.
\end{eqnarray}
Since the smeared neutron structure function is defined as
$\widetilde{F}_2^n \equiv F_2^n/S_n$, we can invert this to
obtain the structure function of a free neutron,
\begin{eqnarray}
\label{F2n}
F_2^n &=& S_n \left( F_2^{D\ {\rm (conv)}} - F_2^p/S_p \right).
\end{eqnarray}

The proton smearing factor, $S_p$, can be computed at each $x$ 
from the function $f_{N/D}(y)$, and a parametrization of the 
$F_2^p$ 
data (for example, the recent fit in Ref.\cite{F2PAR} to the 
combined SLAC, BCDMS and NMC data).
The neutron $F_2^n$ structure function is then derived from
Eq.(\ref{F2n}) taking as a first guess $S_n = S_p$.
These values of $F_2^n$ are then smeared by the function 
$f_{N/D}(y)$,
and the results used to obtain a better estimate for $S_n$.
The new value for $S_n$ is then used in Eq.(\ref{F2n}) to obtain
an improved estimate for $F_2^n$, and the procedure repeated
until convergence is achieved.

\begin{figure}
\centering{\ \psfig{figure=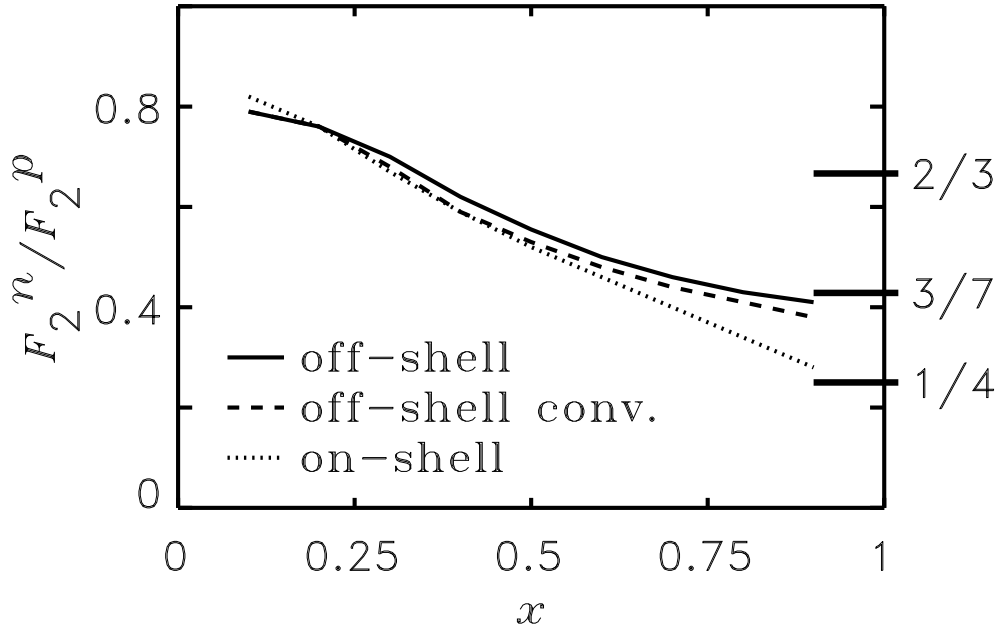,height=10cm}}
\caption{$F_2^n/F_2^p$ ratio as a function of $x$, for the
         off-shell model (solid), off-shell model without
         the convolution-breaking term (dashed),
         and the on-shell model (dotted).
         On the right-hand axis are marked the $x \rightarrow 1$
         limits of the SU(6) symmetric model (2/3), and the predictions
         of the models of Refs.\protect\cite{CLO73,CAR75} (1/4) and
         \protect\cite{FJ,BBS} (3/7).}
\end{figure}

The results of this procedure for $F_2^n/F_2^p$ are presented 
in Fig.2, for both the off-shell calculation (solid) and the 
on-shell model (dotted).
The increase in the off-shell ratio at large $x$ is a direct 
consequence of the deeper trough in the $F_2^D/F_2^N$ ratio 
in Fig.1.
To illustrate the role of the non-convolution correction, 
$\delta^{\rm (off)} F_2^D$, we have also performed the 
analysis setting this term to zero, and approximating 
$F_2^D$ by $F_2^{D\ {\rm (conv)}}(x)$.
The effect of this correction (dashed curve in Fig.2) appears 
minimal.
One can therefore attribute most of the difference between the
off- and on-shell results to kinematics, since both calculations
involve very similar deuteron wave functions.

Before discussing the implications of these results, let us 
briefly outline the connection between structure functions 
at large $x$ and the breaking of SU(6) spin-flavor symmetry.

\subsection{SU(6) Symmetry Breaking}

The large-$x$ region, being valence quark dominated, is where 
SU(6) symmetry breaking effects in valence quark distributions
should be most prominent.
The precise mechanism for the breaking of the spin-flavor
SU(6) symmetry is a basic question in hadronic physics.
In a world of exact SU(6) symmetry, the wave function of a
proton, polarized say in the $+z$ direction, would be simply
\cite{CLO79}:
\begin{eqnarray}
\label{pwfn}
p\uparrow
&=& {1 \over \sqrt{2}}  u\uparrow (ud)_{S=0}\
 +\ {1 \over \sqrt{18}} u\uparrow (ud)_{S=1}\
 -\ {1 \over 3}         u\downarrow (ud)_{S=1}\ \nonumber \\
& &
 -\ {1 \over 3}         d\uparrow (uu)_{S=1}\
 -\ {\sqrt{2} \over 3}  d\downarrow (uu)_{S=1},
\end{eqnarray}
where the subscript $S$ denotes the total spin of the two-quark
component.
In this limit, apart from charge and flavor quantum numbers,
the $u$ and $d$ quarks in the proton would be identical.
The nucleon and $\Delta$ isobar would, for example, be 
degenerate in mass.
In deep-inelastic scattering, exact SU(6) symmetry would 
be manifested in equivalent shapes for the valence quark
distributions of the proton, which would be related simply 
by $u_V(x) = 2 d_V(x)$ for all $x$.
For the neutron to proton structure function ratio this would 
imply:
\begin{eqnarray}
{ F_2^n \over F_2^p }
&=& {2 \over 3}\ \ \ \ \ \ {\rm [SU(6)\ symmetry]}.
\end{eqnarray}

In nature spin-flavor SU(6) symmetry is, of course, broken.
The nucleon and $\Delta$ masses are split by some 300 MeV.
Furthermore, with respect to DIS, it is known that the
$d$ quark distribution is softer
than the $u$ quark distribution, with the neutron/proton 
ratio deviating at large $x$ from the SU(6) expectation.
The correlation between the mass splitting in the {\bf 56} 
baryons and the large-$x$ behavior of $F_2^n/F_2^p$ was 
observed some time ago by Close \cite{CLO73} and Carlitz 
\cite{CAR75}.
Based on phenomenological \cite{CLO73} and Regge \cite{CAR75}
arguments, the breaking of the symmetry in Eq.(\ref{pwfn}) 
was argued to arise from a suppression of the ``diquark'' 
configurations having $S=1$ relative to the $S=0$ configuration, 
namely
\begin{eqnarray}
(qq)_{S=0} &\gg& (qq)_{S=1}, \ \ \ \ \ x \rightarrow 1.
\label{S0dom}
\end{eqnarray}
Such a suppression is in fact quite natural if one observes 
that whatever mechanism leads to the observed $N-\Delta$ 
splitting (e.g. color-magnetic force, instanton-induced 
interaction, pion exchange), it necessarily acts to produce 
a mass splitting between the two possible spin states of the two   
quarks, $(qq)_S$, which act as spectators to the hard collision, 
with the $S=1$ state heavier than the $S=0$ state 
by some 200 MeV \cite{CT}. 
{}From Eq.(\ref{pwfn}), a dominant scalar valence diquark 
component of the proton suggests that in the $x \rightarrow 1$ 
limit $F_2^p$ is essentially given by a single quark distribution 
(i.e. the $u$), in which case:
\begin{eqnarray}
{ F_2^n \over F_2^p }
&\rightarrow& { 1 \over 4 }, \ \ \ \ \
{ d \over u } \rightarrow 0\ \ \ \ \
[S=0\ {\rm dominance}].
\end{eqnarray}
This expectation has, in fact, been built into many
phenomenological fits to the parton distribution data.

An alternative suggestion, based on perturbative QCD,
was originally formulated by Farrar and Jackson \cite{FJ}.
There it was argued that the exchange of longitudinal gluons,
which are the only type permitted when the spins of the two 
quarks in $(qq)_S$ are aligned, would introduce a factor 
$(1-x)^{1/2}$ into the Compton amplitude --- in comparison 
with the exchange of a transverse gluon between quarks with 
spins anti-aligned.
In this approach the relevant component of the proton valence
wave function at large $x$ is that associated with states in
which the total ``diquark'' spin {\em projection}, $S_z$, 
is zero:
\begin{eqnarray}
(qq)_{S_z=0} &\gg& (qq)_{S_z=1}, \ \ \ \ \ x \rightarrow 1.
\end{eqnarray}
Consequently, scattering from a quark polarized in the opposite
direction to the proton polarization is suppressed by a factor
$(1-x)$ relative to the helicity-aligned configuration.

A similar result is also obtained in the treatment of
Brodsky {\em et al.} \cite{BBS} (based on counting-rules),
where the large-$x$ behavior of the parton distribution for 
a quark polarized parallel ($\Delta S_z = 1$) or antiparallel 
($\Delta S_z = 0$) to the proton helicity is given by:
$q^{\uparrow\downarrow}(x) = (1~-~x)^{2n - 1 + \Delta S_z}$,
where $n$ is the minimum number of non-interacting quarks 
(equal to 2 for the valence quark distributions).
In the $x \rightarrow 1$ limit one therefore predicts:
\begin{eqnarray}
{ F_2^n \over F_2^p }
&\rightarrow& {3 \over 7}, \ \ \ \ \
{ d \over u } \rightarrow { 1 \over 5 }\ \ \ \ \
[S_z=0\ {\rm dominance}].
\end{eqnarray}
Note that the $d/u$ ratio {\em does not vanish} in this model.
Clearly, if one is to understand the dynamics of the nucleon's
quark distributions at large $x$, it is imperative that the
consequences of these models be tested experimentally.

The reanalyzed SLAC \cite{WHIT,GOMEZ} data points themselves 
are plotted in Fig.3, at an average value of 
$Q^2 \approx 12$ GeV$^2$.
The very small error bars are testimony to the quality of the 
SLAC $p$ and $D$ data.
The data represented by the open circles have been extracted 
with the on-shell deuteron model of Ref.\cite{FS78}, while the 
filled circles were obtained using the off-shell model of 
Refs.\cite{MSTD,MST}.
Most importantly, the $F_2^n/F_2^p$ points obtained with the
off-shell method appear to approach a value broadly consistent
with the Farrar-Jackson \cite{FJ} and Brodsky {\em et al.} 
\cite{BBS} prediction of 3/7, whereas the data previously 
analyzed in terms of the on-shell formalism produced a ratio 
that tended to the lower value of 1/4.

\begin{figure}
\centering{\ \psfig{figure=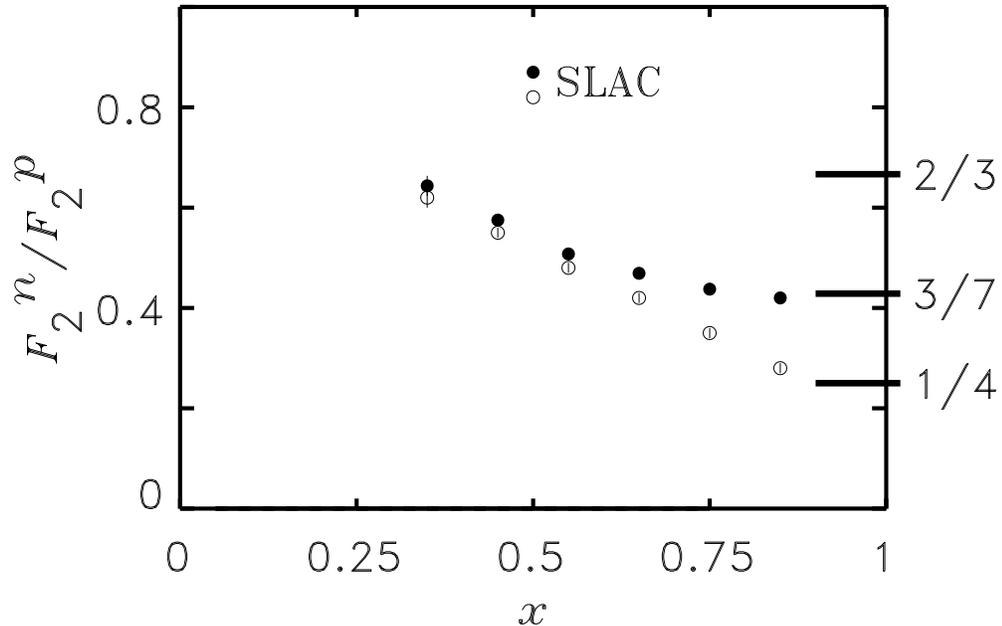,height=10cm}}
\caption{Deconvoluted $F_2^n/F_2^p$ ratio extracted from the
         SLAC $p$ and $D$ data \protect\cite{WHIT,GOMEZ},
         at an average value of $Q^2 \approx 12$
         GeV$^2$, assuming no off-shell effects (open circles),
         and including off-shell effects (full circles).}
\end{figure}

The $d/u$ ratio, shown in Fig.4, is obtained by inverting
$F_2^n/F_2^p$ in the valence quark dominated region.
The points extracted using the off-shell formalism (solid circles)
are again significantly above those obtained previously with the
aid of the on-shell prescription.
In particular, they indicate that the $d/u$ ratio may actually
approach a {\em finite} value in the $x \rightarrow 1$ limit,
contrary to the expectation of the model of Refs.\cite{CLO73,CAR75},
in which $d/u$ tends to zero.
Although it is {\em a priori} not clear at which scale the model
predictions \cite{CLO73,CAR75,FJ,BBS} should be valid, for the 
values of $Q^2$ corresponding to the analyzed data the effects of
$Q^2$ evolution are minimal.

\begin{figure}
\centering{\ \psfig{figure=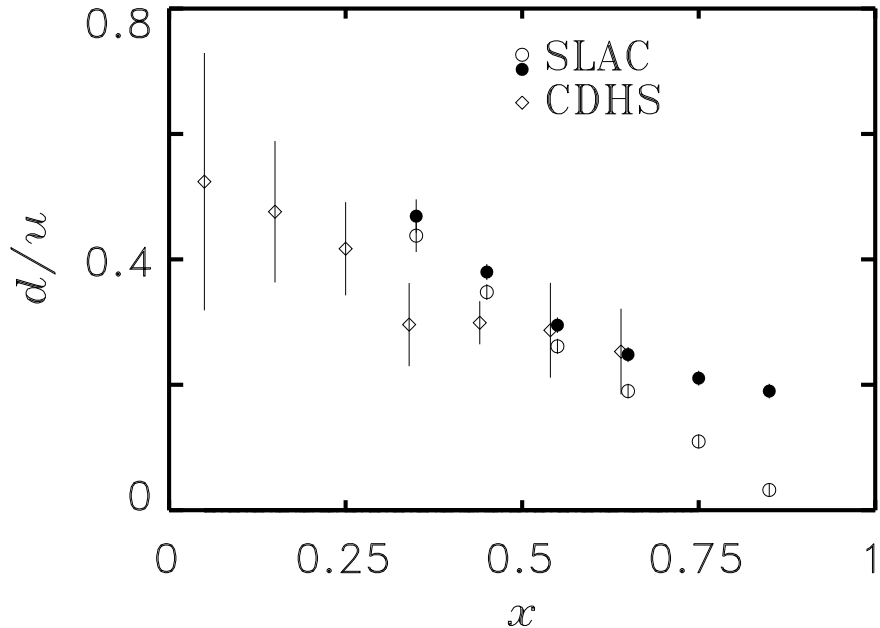,height=10cm}}
\caption{Extracted $d/u$ ratio, using the off-shell deuteron
         calculation (full circles) and using on-shell kinematics
         (open circles).  Also shown for comparison is the ratio
         extracted from neutrino measurements by the CDHS collaboration
         \protect\cite{CDHS}.}
\end{figure}

Naturally it would be preferable to extract $F_2^n$ at large $x$
without having to deal with uncertainties in the nuclear effects.
In principle this could be achieved by using neutrino and
antineutrino beams to measure the $u$ and $d$ distributions in
the proton separately, and reconstructing $F_2^n$ from these.
Unfortunately, as seen in Fig.4, the neutrino data from the
CDHS collaboration \cite{CDHS} do not extend out to very large
$x$ ($x \alt 0.6$), and at present cannot discriminate between
the different methods of analyzing the electron--deuteron data.

The results of our off-shell model are qualitatively similar 
\cite{GOMEZ} to those obtained using the nuclear density method 
suggested by Frankfurt and Strikman \cite{FS88}. 
There the EMC effect in deuterium was assumed to scale with that
in heavier nuclei according to the ratio of the respective nuclear
densities, so that the ratio $F_2^D/F_2^N$ in the trough region 
was depleted by about 4\%, similar to that in Fig.1 (solid curve).
This would give an $F_2^n/F_2^p$ ratio broadly consistent with 3/7.

We should also point out similar consequences for the 
spin-dependent neutron structure function $g_1^n$, where 
the models of Refs.\cite{CLO73,CAR75} and Refs.\cite{FJ,BBS} 
also give different predictions for $g_1^n/g_1^p$ as 
$x \rightarrow 1$, namely 1/4 and 3/7, respectively.
Quite interestingly, while the ratio of polarized to 
unpolarized $u$ quark distribution is predicted to be 
the same in the two models,
\begin{eqnarray}
{ \Delta u \over u }
&\rightarrow& 1\ \ \ \ [S=0\ or\ S_z=0\ {\rm dominance}],
\end{eqnarray}
the results for the $d$-quark distribution ratio differ even
in sign:
\begin{mathletters}
\begin{eqnarray}
{ \Delta d \over d }
&\rightarrow& - {1 \over 3}\ \ \ \ [S=0\ {\rm dominance}],\\
&\rightarrow& 1\ \ \ \ [S_z=0\ {\rm dominance}].
\end{eqnarray}
\end{mathletters}%
To extract information on the polarized parton densities at 
large $x$ that is capable of discriminating between these 
predictions, the same care will need to be taken when subtracting 
the nuclear effects from $g_1^D$ and $g_1^{^3He}$.
In particular, the results of Refs.\cite{KMPW,MPT} indicate that 
while the simple prescription \cite{DPOL} of subtracting the 
$g_1^p$ structure function from the $D$ data, modified only by 
the deuteron $D$-state probability, is surprisingly good for 
$x \alt 0.6$, it is completely inadequate for $x \agt 0.7$.

Having seen that how one handles nuclear corrections can 
critically affect the deciphering of the physical implications 
of the extracted structure function at large $x$, we now examine 
the consequences of nuclear effects in deuterium in the small-$x$
region.

\section{Small $x$}

At small values of $x$ the impulse approximation should 
eventually break down.  
Indeed, one finds that coherent multiple scattering effects 
become very important when the characteristic time scale $1/Mx$ 
of the DIS process becomes larger than the typical average 
distance between bound nucleons in the nucleus, which occurs
typically for $x \alt 0.1$.
These effects are seen, for example, in the low-$x$ depletion 
of the nuclear EMC ratio, $F_2^A/F_2^D$. 
Any shadowing in the deuteron itself should therefore produce
a depletion in the $F_2^D/F_2^N$ ratio at small $x$ 
\cite{KWBD,KW,BK,MTD,MTA,OTHERD}.

\subsection{Nuclear Shadowing}

The rescattering of the virtual photon from several nucleons 
in a nucleus is usually described within the non-relativistic 
Glauber multiple scattering formalism.
Relativistic corrections will amount to a few percent out of 
a shadowing correction to $F_2^D$ that will itself be a few 
percent in total, and hence can be neglected.
For the deuteron the only contribution in the Glauber series 
comes from the double scattering process.

At small $x$, nuclear binding and Fermi motion corrections  
can be neglected, and the total deuteron structure function 
written as:
\begin{eqnarray}
F_2^D(x,Q^2)
&\approx& F_2^p(x,Q^2) + F_2^n(x,Q^2) 
+ \delta^{\rm (shad)} F_2^D(x,Q^2).  
\end{eqnarray}
In modeling the shadowing correction, $\delta^{\rm (shad)} F_2^D$,
our approach is to take a two-phase model, similar to that of 
Kwiecinski and Badelek \cite{KWBD,KW,BK}.
At high virtuality the interaction of the virtual photon with 
the nucleus is parametrized in terms of diffractive scattering 
through the double and triple Pomeron, as well as scattering 
from exchanged mesons in the deuteron.
On the other hand, at low virtuality it is most natural to apply
a vector meson dominance (VMD) model, in which the virtual photon
interacts with the nucleons via its hadronic structure, namely the 
$\rho^0$, $\omega$ and $\phi$ mesons.
The latter contribution vanishes at sufficiently high $Q^2$, but  
for $Q^2 \alt 1$ GeV$^2$ it is in fact responsible for the majority 
of the $Q^2$ variation.

For the diffractive component, Pomeron ($I\!\!P$) exchange 
between the projectile and two or more constituent nucleons 
models the interaction of partons from different nucleons 
within the deuteron.
Assuming factorization of the diffractive cross section, the 
shadowing correction (per nucleon) to the deuteron structure
function $F_2^D$ from $I\!\!P$-exchange is written as a 
convolution of the Pomeron structure function, $F_2^{I\!P}$, 
with a distribution function (``flux factor''), $f_{I\!P/D}$, 
describing the number density of exchanged Pomerons:
\begin{eqnarray}
\label{dFAP}
\delta^{(I\!P)} F_2^D(x,Q^2)
&=& \int_{y_{min}}^2\ dy\ 
f_{I\!P/D}(y)\ F_2^{I\!P}(x_{I\!P},Q^2),
\end{eqnarray}
where, within the non-relativistic approximation
\cite{KWBD,KW,BK,MTD,MTA}, 
\begin{eqnarray}
f_{I\!P/D}(y)
&=& - \frac{\sigma_{pp}}{8 \pi^2}\ \frac{1}{y}
\int d^2{\bf k}_T\ S_{D}({\bf k}^2).
\end{eqnarray}
Here $y = x (1+M_X^2/Q^2)$ the light-cone momentum fraction
carried by the Pomeron ($M_X$ is the mass of the diffractive
hadronic debris), and $x_{I\!P} = x/y$ is the momentum fraction
of the Pomeron carried by the struck quark in the Pomeron.
The deuteron form factor, $S_{D}({\bf k}^2)$, is given in 
terms of the coordinate space wave functions:
\begin{eqnarray}
S_{D}({\bf k}^2)
&=& \int_{0}^{\infty} dr \left( u^{2}(r) + w^{2}(r) \right)  
j_{0}(|{\bf k}| r),
\end{eqnarray}
where $j_{0}$ is the spherical Bessel function.
Within experimental errors, the factorization hypothesis, 
as well as the $y$ dependence of $f_{I\!P/A}(y)$
\cite{KWBD,KW,BK,MTD,MTA}, are consistent with the recent 
HERA data \cite{HERA} obtained from observations 
of large rapidity gap events in diffractive $ep$ scattering.
These data also confirm previous findings that the Pomeron 
structure function contains both a hard and a soft component:\
$ F_2^{I\!P}(x_{I\!P},Q^2)
= F_2^{I\!P ({\rm hard})}(x_{I\!P},Q^2)
+ F_2^{I\!P ({\rm soft})}(x_{I\!P},Q^2)$.
The hard component of $F_2^{I\!P}$ is generated from
an explicit $q\bar q$ component of the Pomeron, and has
an $x_{I\!P}$ dependence given by\
$x_{I\!P} (1-x_{I\!P})$ \cite{DOLA}, in agreement
with the recent diffractive data \cite{HERA}.
The soft part, which is driven at small $x_{I\!P}$ by the
triple-Pomeron interaction \cite{KWBD}, has a sea quark-like
$x_{I\!P}$ dependence, with normalization fixed by the 
triple-Pomeron coupling constant.

The dependence of $F_2^{I\!P}$ on $Q^2$ at large $Q^2$ arises
from radiative corrections to the parton distributions in the
Pomeron \cite{KW}, which leads to a weak, logarithmic, $Q^2$ 
dependence for the shadowing correction $\delta^{(I\!P)} F_2^D$.
The low-$Q^2$ extrapolation of the $q\bar q$ component is
parametrized by applying a factor $Q^2/(Q^2 + Q_0^2)$, where
$Q_0^2 \approx 0.485$ GeV$^2$ \cite{DLQ} may be interpreted
as the inverse size of partons inside the virtual photon.
For the nucleon sea quark densities relevant for
$F_2^{I\!P ({\rm soft})}$
we use the recent parametrization from Ref.\cite{DLQ}, which 
includes a low-$Q^2$ limit consistent with the real photon data, 
in which case the total Pomeron contribution
$\delta^{(I\!P)} F_2^D \rightarrow 0$ as $Q^2 \rightarrow 0$.

To adequately describe shadowing for small $Q^2$ requires one to
use a higher-twist mechanism, such as vector meson dominance.
VMD is empirically based on the observation that some aspects
of the interaction of photons with hadronic systems resemble
purely hadronic interactions.
In terms of QCD this is understood in terms of a coupling of the
photon to a correlated $q\bar q$ pair of low invariant mass,
which may be approximated as a virtual vector meson.
One can then estimate the amount of shadowing in terms of the
multiple scattering of the vector meson using Glauber theory.
The corresponding correction (per nucleon) to the nuclear
structure function is:
\begin{eqnarray}
\label{dFAV}
\delta^{(V)} F_2^D(x,Q^2)
&=& \frac{ Q^{2} }{ \pi }
\sum_V
{ \delta\sigma_{VD} \over f_V^2 (1 + Q^2/M_V^2)^2 },
\end{eqnarray}
where 
\begin{eqnarray}
\delta \sigma_{VD}
&=& -\frac{ \sigma_{VN}^{2} }{ 8 \pi^{2} }
\int d^2{\bf k}_T S_{D}({\bf k}^2)    
\end{eqnarray}
is the shadowing correction to the vector meson---nucleus 
cross section, $f_V$ is the photon---vector meson coupling 
strength, and $M_V$ is the vector meson mass.

In practice, only the lowest mass vector mesons
($V = \rho^0, \omega, \phi$) are important at low $Q^2$.
For $Q^2 \rightarrow 0$ and fixed $x$, $\delta^{(V)} F_2^D$
disappears due to the vanishing of the total $F_2^D$.
Furthermore, since this is a higher twist effect, shadowing 
in the VMD model dies off quite rapidly between $Q^2 \sim 1$ 
and 10 GeV$^2$, so that for $Q^2 \agt 10$ GeV$^2$ it is 
almost negligible --- leaving only the diffractive term, 
$\delta^{(I\!P)} F_2^D$.
(Note that at fixed $\nu$, for decreasing $Q^2$ the ratio
$F_2^D/F_2^p$ approaches the photoproduction limit.)

Another potential source of shadowing arises from the exchange 
of mesons between the nucleons and the probe.  
It has previously been suggested \cite{KAP} that this leads to 
some {\em antishadowing} corrections to $F_2^D(x)$.
The total contribution to the deuteron structure function from 
meson exchange is:  
\begin{eqnarray}
\delta^{(M)} F_2^D(x,Q^2)
&=& \sum_M
\int_x dy\ f_{M/D}(y)\ F_2^M(x/y,Q^2), 
\end{eqnarray}
where $M = \pi, \rho, \omega, \sigma$.
The virtual meson structure function, $F_2^M$, one approximates
by the (real) pion structure function, data for which exist
from Drell-Yan production.
The exchange-meson distribution functions $f_M(y)$ are obtained   
from the non-relativistic reduction of the nucleon---meson 
interaction given in Refs.\cite{MTD,KAP}.
In practice pion exchange is the dominant process, and this 
gives a positive contribution to $\delta^{(M)} F_2^D(x,Q^2)$.
The exchange of the fictitious $\sigma$ meson (which represents
correlated $2 \pi$ exchange) also gives rise to antishadowing 
for small $x$.
Vector mesons ($\rho$, $\omega$) exchange cancels some of this
antishadowing, however the magnitude of these contributions is 
smaller.
In fact, for soft meson--nucleon vertices 
($\Lambda_M \alt 1.3$ GeV) all contributions other
than that of the pion are totally negligible.

\subsection{Neutron Structure Function at Small $x$}

For $x \alt 0.1$ the magnitude of the (negative) Pomeron/VMD 
shadowing is larger than the (positive) meson-exchange 
contribution, so that the total $\delta^{\rm (shad)} F_2^D$ 
is negative. 
For larger $x$ ($\approx 0.1-0.2$) there is a very small 
amount of antishadowing, which is due mainly to the VMD 
contribution, and also to the pion-exchange contribution.
For the NMC kinematics ($x > 0.004$, $Q^2=4$ GeV$^2$) 
\cite{N_D}, the overall effect on the shape of the 
neutron structure function is a $1 - 2 \%$ {\em increase}
in $F_2^n/F_2^{n, \rm bound}$ for $x \alt 0.01$, where 
$F_2^{n, \rm bound} \equiv F_2^D - F_2^p$.
The presence of shadowing in the deuteron would be confirmed 
through observation of a deviation from unity in the 
$F_2^D/F_2^p$ structure function ratio in the kinematic 
region where Regge theory is expected to be valid.
Although the exact value of $x$ below which the proton and
(free) neutron structure functions become equivalent is not
known, it is expected that at low enough $x$,
$F_2^p \rightarrow F_2^n$, in which case
$F_2^D/F_2^p \rightarrow 1 + \delta^{\rm (shad)} F_2^D/F_2^p$.
While for the lowest NMC data point it may be debatable 
whether the Regge region is reached, the E665 Collaboration 
\cite{E_D} has taken data to very low $x$, $x \sim 10^{-5}$,
which should be much nearer the onset of Regge behavior.

\begin{figure}
\centering{\ \psfig{figure=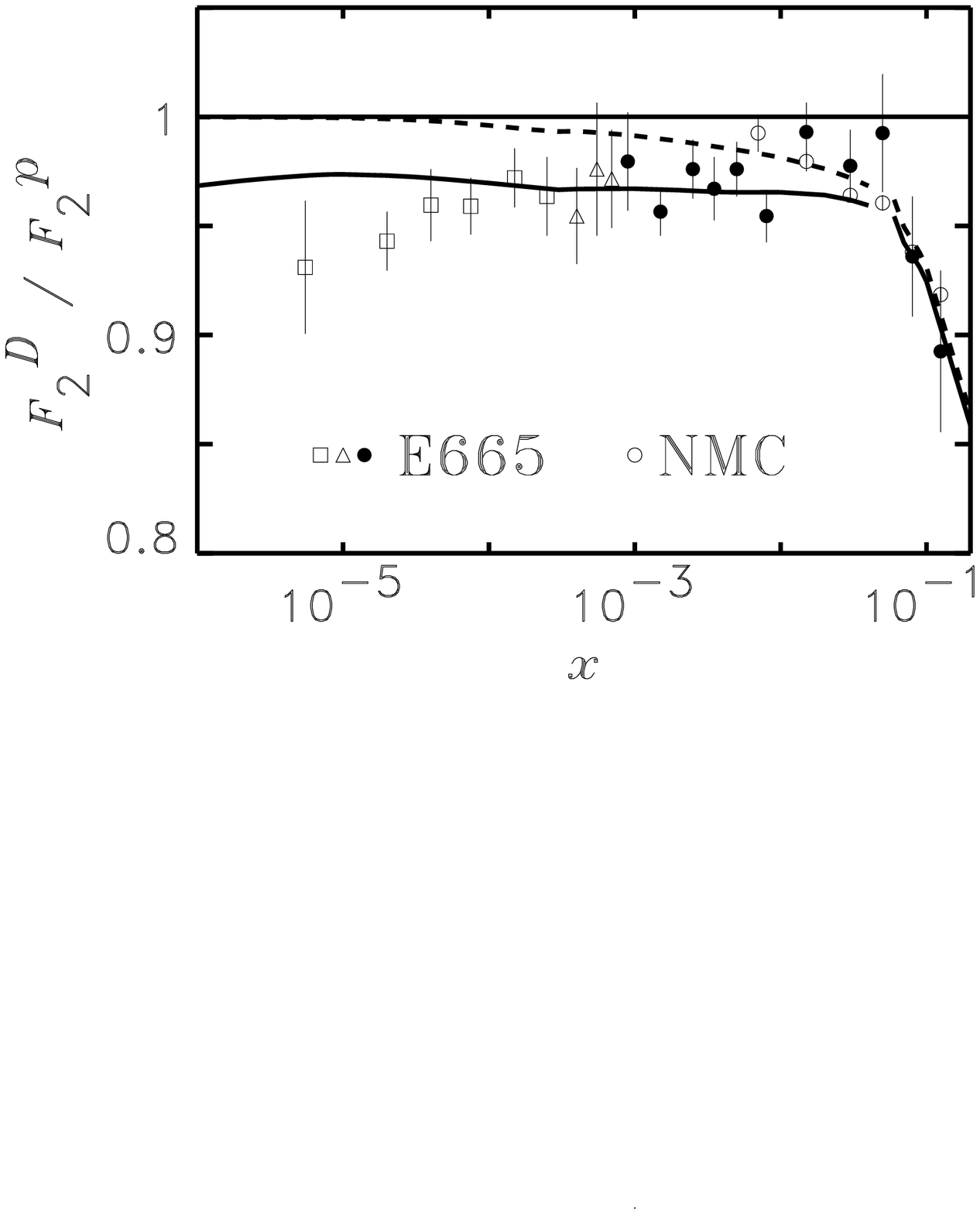,height=10cm}}
\caption{$x$ dependence of the $D/p$ structure function ratio,
        compared with the low-$x$ E665 data \protect\cite{E_D}
        and NMC data \protect\cite{N_D} at larger $x$.
        The dashed curve is the result without any shadowing
        correction.}
\end{figure}

In Fig.5 we show the low $x$ E665 Collaboration data \cite{E_D},
as well as the earlier NMC data at larger $x$ \cite{N_D}
(note that the E665  data are not taken at fixed $Q^2$).
The calculated ratio with a small shadowing correction is shown
by the solid curve, while the result for the case of no shadowing
is indicated by the dashed curve.
Assuming $F_2^p = F_2^n$ in the lowest $x$-bins, the data clearly 
favor the shadowing scenario.

\subsection{Are 1\% Effects Worth Worrying About? 
	 	-- Gottfried Sum Rule}  

The main interest in the NMC measurment of the neutron structure
function at low $x$ was to test accurately the Gottfried sum rule,
\begin{mathletters}
\begin{eqnarray}
S_G
&=& \int_0^1 dx\ \frac{ F_2^p(x) - F_2^n(x) }{ x } \\
&=& {1 \over 3} - \int_0^1 dx\ (\bar d(x) - \bar u(x))
\end{eqnarray}
\end{mathletters}%
which, in the naive quark model where $\bar d = \bar u$,
is $S_G^{\rm qm} = 1/3$. 
Ignoring nuclear effects, the experimental value obtained by 
the NMC was $S_G^{\rm exp} = 0.258 \pm 0.017$, indicating a 
violation of SU(2) flavor symmetry in the proton's sea.
However, because the structure function difference in $S_G$ 
is weighted by a factor $1/x$, any small differences between 
$F_2^n$ and $F_2^{n, \rm bound}$ could be amplified for 
$x \rightarrow 0$.

Including the above shadowing correction, the overall effect 
on the experimental value for $S_G$, 
\begin{eqnarray}
S_G
&=& S_G^{\rm exp}
 + \int_0^1 dx { \delta^{\rm (shad)} F_2^D(x) \over x },   
\end{eqnarray}
is a reduction of between --0.010 and --0.026, or about 
4 and 10\% of the measured value without putting in the 
shadowing correction \cite{MTD}. 
Therefore a value that reflects the ``true'' Gottfried sum 
should be around $S_G \approx 0.2$.
This is some 30\% reduction from the naive quark model 
prediction, $S_G^{\rm qm}$.  
Hence the tiny shadowing effect in the deuteron results 
in a more significant violation of flavor SU(2) symmetry 
in the proton sea.

\section{Summary}

In summary, we have reanalyzed the latest proton and deuteron 
structure function data at small and large $x$, in order to   
obtain more reliable information on the structure of the neutron  
in the $x \rightarrow 0$ and $x \rightarrow 1$ limits.

At small $x$, we have estimated the nuclear shadowing arising 
from the double scattering of the virtual photon from both nucleons 
in the deuteron.
To cover the entire range of $Q^2$ accessible to current 
experiments, the $\gamma^* N$ interaction is described in terms
of the VMD model, which models the correlated $q\bar q$ pair 
excitations of the virtual photon, together with Pomeron exchange 
for the uncorrelated $q\bar q$ pair interactions with the nucleon.
Such a hybrid model is particularly necessary if one is to 
reproduce the observed $Q^2$ dependence of the EMC ratio at 
small $x$ \cite{MTQ}.
In addition, we have also included contributions from the 
exchange of mesons, which effectively cancels as much as half 
of the shadowing from the VMD/Pomeron-exchange mechanisms alone. 
The net effect is a $\alt 1\%$ reduction of $F_2^D$ for 
$x \sim 0.004$, or equivalently a $\alt 2\%$ increase in 
the neutron structure function over the uncorrected $F_2^n$.
This is consistent with the recent measurement by the Fermilab
E665 Collaboration \cite{E_D} of the ratio $F_2^D/F_2^p$ down  
to $x \sim 10^{-5}$, which deviates from unity by about 4\%,
and suggests the presence of shadowing in the deuteron.
Although the absolute values of the shadowing corrections to 
$F_2^D$ are small, because they are concentrated at small $x$, 
the effect on the Gottfried sum $S_G$ is a further reduction 
of up to $\sim 10\%$ over the value measured by NMC \cite{N_D}.

Including all of the currently known nuclear effects in the 
deuteron at large $x$, namely Fermi motion, binding, and nucleon
off-mass-shell effects, we find that the total EMC effect 
is $\sim 2\%$ larger than in previous calculations based 
on on-mass-shell kinematics, from which binding effects were 
omitted.
The larger deviation from unity for $0.5 \alt x \alt 0.8$ 
in $F_2^D/F_2^N$ translates into an increase in the ratio 
$F_2^n/F_2^p$.
Our results indicate that as $x \rightarrow 1$ the limiting 
value of $F_2^n/F_2^p$ is above the previously accepted result 
of 1/4, and broadly consistent with the perturbative QCD 
expectation of 3/7.  
This also implies that the $d/u$ ratio approaches a 
{\em non-zero} value of 1/5 as $x \rightarrow 1$.

Finally, for definitive tests of the nuclear effects in the 
deuteron one would like model-independent information on the 
neutron at both low and high $x$. 
In principle, this can be achieved with high-precision data 
from neutrino-proton experiments, from which individual flavor 
distributions can be determined,
and the neutron structure function inferred from charge symmetry.
Unfortunately, both the statistics on the neutrino data and the 
coverage in $x$ will not allow this in the near future.   
It has therefore been suggested that one might perform a series  
of semi-inclusive experiments on deuteron targets, measuring in 
coincidence both the scattered lepton and recoiling proton or 
neutron.
This would help constrain the deuteron wave function over a large
range of kinematics, and hence enable the $y$-dependence of the 
nucleon momentum distribution functions to be mapped out directly.
Such experiments are already planned for CEBAF and HERMES 
\cite{SEMI}, and should provide critical information on the 
size and importance of relativistic and other short-distance 
nuclear phenomena in the deuteron.

\acknowledgements

We would like to thank S.A.Kulagin, G.Piller, A.W.Schreiber 
and W.Weise for their contributions to the issues discussed 
here.
W.M. thanks the organizers of the IInd. Cracow Epiphany 
Conference on Proton Structure for their hospitality, and 
the H. Niewodnicza\'nski Institute of Nuclear Physics for 
support during the stay in Cracow.  
This research was partially supported by the Australian 
Research Council and the U.S. Department of Energy grant 
\# DE-FG02-93ER-40762.



\references

\bibitem{CONV}
R.L.Jaffe,
in {\em Relativistic Dynamics and
Quark-Nuclear Physics},
eds. M.B.Johnson and A.Pickleseimer
(Wiley, New York, 1985);
%
S.V.Akulinichev, S.A.Kulagin and G.M.Vagradov,
Phys.Lett. B 158 (1985) 485;
%
G.V.Dunne and A.W.Thomas,
Nucl.Phys. A455 (1986) 701;
%
H.Jung and G.A.Miller,
Phys.Lett. B 200 (1988) 351;
%
F.Gross and S.Liuti,
Phys.Rev. C 45 (1992) 1374;
%
S.A.Kulagin, G.Piller and W.Weise,
Phys.Rev. C 50 (1994) 1154.

\bibitem{KMPW}
S.Kulagin, W.Melnitchouk, G.Piller and W.Weise,
Phys.Rev. C 52 (1995) 932.

\bibitem{DREL}
W.W.Buck and F.Gross,
Phys.Rev. D 20 (1979) 2361;
%
J.A.Tjon,
Nucl.Phys. A 463 (1987) 157C;
%
D.Plumper and M.F.Gari,
Z.Phys. A 343 (1992) 343;
%
F.Gross, J.W.Van Orden and K.Holinde,
Phys.Rev. C 45 (1992) 2094.

\bibitem{MSTD}
W.Melnitchouk, A.W.Schreiber and A.W.Thomas,
Phys.Lett. B 335 (1994) 11.

\bibitem{MST}
W.Melnitchouk, A.W.Schreiber and A.W.Thomas,
Phys.Rev. D 49 (1994) 1183.

\bibitem{FS78}
L.L.Frankfurt and M.I.Strikman,
Phys.Lett. 76 B (1978) 333;
Phys.Rep. 76 (1981) 215.

\bibitem{WHIT}
L.W.Whitlow {\em et al.},
Phys.Lett. B 282 (1992) 475.

\bibitem{EMC}
EM Collaboration, J.J.Aubert {\em et al.},
Nucl.Phys. B293 (1987) 740.

\bibitem{US}
T.Uchiyama and K.Saito,
Phys.Rev. C 38 (1988) 2245.  

\bibitem{KU}
L.P.Kaptari and A.Yu.Umnikov,
Phys.Lett. B 259 (1991) 155.

\bibitem{BT}
M.A.Braun and M.V.Tokarev,
Phys.Lett. B 320 (1994) 381.

\bibitem{BODEK}
A.Bodek {\em et al.},
Phys.Rev. D 20 (1979) 1471;
%
A.Bodek and J.L.Ritchie,
Phys.Rev. D 23 (1981) 1070.

\bibitem{UKK}
A.Yu.Umnikov, F.C.Khanna and L.P.Kaptari,
Z.Phys. A 348 (1994) 211.

\bibitem{VATO}
A.H\"ocker and V.Kartvelishvili,
Manchester preprint MC-TH-95/15, LAL-95/55 (1995);
%
V.Blobel, 
DESY preprint DESY-84-118 (1984).

\bibitem{F2PAR}
NM Collaboration, M.Arneodo {\em et al.}, 
Preprint CERN-PPE/95-138 (1995).

\bibitem{CLO79}
F.E.Close,
{\em An Introduction to Quarks and Partons}
(Academic Press, 1979).

\bibitem{CLO73}
F.E.Close,
Phys.Lett. 43 B (1973) 422.

\bibitem{CAR75}
R.Carlitz, 
Phys.Lett. 58 B (1975) 345.

\bibitem{CT}
F.E.Close and A.W.Thomas,
Phys.Lett. B 212 (1988) 227.

\bibitem{FJ}
G.R.Farrar and D.R.Jackson,
Phys.Rev.Lett. 35 (1975) 1416.

\bibitem{BBS}
S.J.Brodsky, M.Burkardt and I.Schmidt,
Nucl.Phys. B441 (1995) 197.

\bibitem{GOMEZ}
J.Gomez {\em et al.},
Phys.Rev. D 49 (1994) 4348.

\bibitem{CDHS}
H.Abramowicz {\em et al.},
Z.Phys. C 25 (1983) 29.

\bibitem{FS88}
L.L.Frankfurt and M.I.Strikman,
Phys.Rep. 160 (1988) 235.

\bibitem{MPT}
W.Melnitchouk, G.Piller and A.W.Thomas,
Phys.Lett. B 346 (1995) 165.  

\bibitem{DPOL}
SM Collaboration, D.Adams {\em et al.},
Phys. Lett. B 357 (1995) 248;
%
E143 Collaboration, K.Abe {\em et al.},
Phys.Rev.Lett. 75 (1995) 25.

\bibitem{KWBD}
J.Kwiecinski and B.Badelek,
Phys.Lett. B 208 (1988) 508.

\bibitem{KW}
J.Kwiecinski,
Z.Phys. C 45 (1990) 461.

\bibitem{BK}
B.Badelek and J.Kwiecinski,
Nucl.Phys. B370 (1992) 278;
Phys.Rev. D 50 (1994) 4.

\bibitem{MTD}
W.Melnitchouk and A.W.Thomas,
Phys.Rev. D 47 (1993) 3783.

\bibitem{MTA}
W.Melnitchouk and A.W.Thomas,
Phys.Lett. B 317 (1993) 437.

\bibitem{OTHERD}
V.R.Zoller,
Z.Phys. C 54 (1992) 425;
%
G.Piller, W.Ratzka and W.Weise,
Z.Phys. A (1995), in print;
%
H.Khan and P.Hoodbhoy,
Phys.Lett. B 298 (1993) 181.

\bibitem{HERA}
H1 Collaboration, T.Ahmed {\em et al.},
Phys.Lett. B 348 (1995) 681.

\bibitem{DOLA}  
A.Donnachie and P.V.Landshoff,
Phys.Lett. B 191 (1987) 309.

\bibitem{DLQ}   
A.Donnachie and P.V.Landshoff,
Z.Phys. C 61 (1994) 139.

\bibitem{KAP} 
L.P.Kaptari, A.I.Titov, E.L.Bratkovskaya and A.Yu.Umnikov,
Nucl.Phys. A 512 (1990) 684.

\bibitem{N_D}
NM Collaboration, M.Arneodo {\em et al.},
Phys.Rev. D 50 (1994) 1.

\bibitem{E_D}
E665 Collaboration, M.R.Adams {\em et al.},
Phys.Rev.Lett. 75 (1995) 1466;
Phys.Lett. B 309 (1993) 477.

\bibitem{MTQ}
W.Melnitchouk and A.W.Thomas,
Phys.Rev. C 52 (1995) 3373.

\bibitem{SEMI}
S.Kuhn {\em et al.}, CEBAF proposal PR-94-102.
\end{document}